\documentclass[11pt]{article}

\usepackage[english]{babel}
\usepackage[utf8]{inputenc}
\usepackage{johd}
\usepackage{fancyhdr}
\usepackage{amsmath}
\usepackage{mathrsfs}
\usepackage[version=3]{mhchem} 
\usepackage{graphicx}
\usepackage{dcolumn}
\usepackage{bm}
\usepackage{hyperref}
\hypersetup{
    colorlinks=true,
    linkcolor=blue,
    citecolor=blue,
}
\usepackage[utf8]{inputenc}
\usepackage[T1]{fontenc}
\usepackage{mathptmx}
\usepackage{etoolbox}
\usepackage{xcolor, soul}

\newcommand{\blue}[1]{#1}

\newcommand{\fullFRAC}[2]{\frac{\mathrm{d}#1}{\mathrm{d}#2}}

\allowdisplaybreaks

\title{\textbf{Can 2D materials enable passively $Q$-switched lasers in the nanoscale?}}

\author{Thomas Christopoulos$^{a,b,*,\dagger}$, Johanne Hizanidis$^{c,d}$, Georgios Nousios$^{b,e}$, \\
        Emmanouil E. Kriezis$^{b}$, and Odysseas Tsilipakos$^{a,*}$ \\
        \small $^{a}$Theoretical and Physical Chemistry Institute, \\ 
        \small       National Hellenic Research Foundation, Athens 11635, Greece. \\
        \small $^{b}$School of Electrical and Computer Engineering, \\
        \small       Aristotle University of Thessaloniki, Thessaloniki 54124, Greece. \\
        \small $^{c}$Institute of Electronic Structure and Laser,\\
        \small       Foundation for Research and Technology-Hellas, Herakleio, Crete 70013, Greece. \\
        \small $^{d}$Institute of Nanoscience and Nanotechnology, \\
        \small       National Center for Scientific Research ``Demokritos'', Athens 15341, Greece. \\
        \small $^{e}$Istituto per la Microelettronica e Microsistemi,\\
        \small       Consiglio Nazionale delle Ricerche, Roma 00133, Italy. \\
        \small $^{*}$Corresponding authors: \tt{cthomasa@eie.gr}; \tt{otsilipakos@eie.gr} \\
        \small $^{\dagger}$Current address: Laboratoire Photonique, Num\'erique et Nanosciences (LP2N), \\
        \small       IOGS-University of Bordeaux-CNRS, 33400 Talence, France.
}

\date{} 

\begin{document}

\maketitle
\thispagestyle{empty}

\begin{abstract} 
    Achieving compact on-chip pulsed lasers with attractive performance metrics and compatibility with the silicon photonics platform is an important, yet elusive, goal in contemporary nanophotonics. Here, the fundamental question of whether 2D materials can be utilized as both gain and saturable absorption media to enable compact integrated passively $Q$-switched nanophotonic lasers is posed and addressed by examining a broad range of 2D material families. The study is conducted by developing a temporal coupled-mode theory framework involving semi-classical rate equations that is capable of rigorously handling gain and saturable absorption by 2D materials, allowing to perform stability and bifurcation analysis covering broad parameter spaces. The range of pulse-train metrics (repetition rate, pulse width, peak power) that can be obtained via different 2D materials is thoroughly assessed. Our work illustrates that nanophotonic cavities enhanced with 2D materials can enable passive $Q$-switching with repetition rates ranging up to 50~GHz, short pulse duration down to few picoseconds, and peak power exceeding several milliwatts. Such attractive metrics, along with the ultrathin nature of 2D materials and the ability to electrically tune their properties, demonstrate the potential of the proposed platform for compact and flexible integrated laser sources.
\end{abstract}


\clearpage
\section{\label{sec:Intro}Introduction}

    $Q$-switching is one of the main techniques to achieve pulsed lasing with short pulse duration, high peak power, and controlled repetition rate \cite{Siegman,SalehBook,KellerBook}. It is based on switching the losses (quality factor, $Q$) of a resonant cavity between high and low levels, in order to suppress/enable light emission. This can be achieved either by actively modulating cavity loss (active $Q$-switching) or by utilizing materials that passively suppress their optical losses (passive $Q$-switching) through, e.g., saturable absorption (SA). Active $Q$-switching allows for more flexibility and was the first to be demonstrated \cite{KellerBook}; however, passive $Q$-switching has attracted significant interest due to its simplicity and has been demonstrated in various platforms over the last 30 years \cite{Siegman}.

    Passive $Q$-switching can be achieved in any laser system that additionally incorporates a saturable absorber; it has been demonstrated in external cavity lasers \cite{Sphler1999,Butler2012,Zhang2019,Hao2020,Li2020b}, fiber lasers \cite{Luo2010,Liu2011a,Popa2011,Tiu2019,Yamashita2019,Zhang2020}, and, more recently, integrated cavities \cite{Charlet2011,Yu2017b,Hou2018,Shtyrkova2019}. The emerging pulse-train characteristics can differ vastly  between the aforementioned platforms in terms of pulse duration, peak power, and repetition rate. In general, larger cavities (cavities with long roundtrip times) lead to longer pulses with higher energies, whereas smaller systems achieve shorter pulses at the expense of lower peak power \cite{Siegman,SalehBook,KellerBook}. Currently, an important yet elusive goal remains to demonstrate nanoscale integrated pulsed lasers compatible with the silicon photonics platform. It is also desirable to be able to cover a wide range of pulse-train characteristics, in order to satisfy different applications. 
    
    Recently, 2D materials have been receiving increased interest for laser sources due to their attractive optoelectronic properties.  They were first used as saturable absorbers for passive $Q$-switching in fiber lasers \cite{Tiu2019,Yamashita2019,Zhang2020}. In parallel, 2D materials have been also investigated for providing gain \cite{Wu2015,Ye2015,Ceballos2015,Miller2017,Karni2019,Paik2019}. Only very recently, an integrated laser where 2D materials are used for \textit{both} gain and SA has been proposed \cite{Nousios2024}. Here, we attempt to address the important question of whether 2D materials can efficiently perform both operations and enable passively $Q$-switched lasers in the nanoscale. Importantly, this would allow for ultracompact implementations (2D materials are ultrathin) and the ability to electrically tune the emission and SA properties, leading to additional flexibility. We will explore different 2D material families (e.g., graphene, transition metal dichalcogenides, MX-enes, topological insulators) aiming to assess the range of pulse-train characteristics that can be achieved with these novel ultrathin materials. 
    
    For the analysis, we utilize rigorous numerical tools to model the lasing process and examine its output characteristics. Specifically, the lasing process including the loss saturation mechanism is efficiently described by a system of first-order coupled ordinary differential equations. They are derived based on temporal coupled-mode theory (CMT) \cite{HausBook,Christopoulos2024Tut} and semi-classical rate equations describing the carrier evolution on the gain and SA media \cite{Chua2011,Ataloglou2018,Benzaouia2022,Nousios2023,Nousios2024,Christopoulos2024Tut,Hizanidis2024,Hizanidis2024arXiv}.  Additionally, the performance of the examined configurations is assessed by utilizing stability and bifurcation analysis tools, as well as a powerful continuation algorithm, to cover quickly and efficiently large parameter spaces \cite{Dhooge2003,Guckenheimer2002}. Similar tools have been used to capture the rich dynamics of coupled systems with or without gain. \cite{Terrien2018,Zhiyenbayev2019,Kominis2020,Valagiannopoulos2021,Benzaouia2022}
    
    The rest of the paper is organized as follows. In section~\ref{sec:Framework} we briefly present the mathematical model that we use for the description of the considered cavities and discuss the continuation, stability, and bifurcation analysis tools. Section~\ref{sec:ExamplesBifurcation} includes a brief presentation of the typical experimentally-specified electromagnetic properties of various 2D materials. The ability to obtain pulsed output ($Q$-switching) with different material combinations is thoroughly discussed. Subsequently, in section~\ref{sec:ExamplPulses} some promising 2D materials are selected to examine the possible range of pulse-train characteristics (repetition rate, pulse width, peak power) that become accessible with the proposed platform. Finally, in section~\ref{sec:Concl} we discuss how these metrics compare with those of other approaches for passive $Q$-switching, highlighting the potential of integrated nanophotonic cavities enhanced with 2D materials as nanoscale pulsed lasers.

\section{\label{sec:Framework}Coupled-mode theory and stability analysis of nanophotonic laser cavities}

\begin{figure}[!t]
        \centering
        \includegraphics{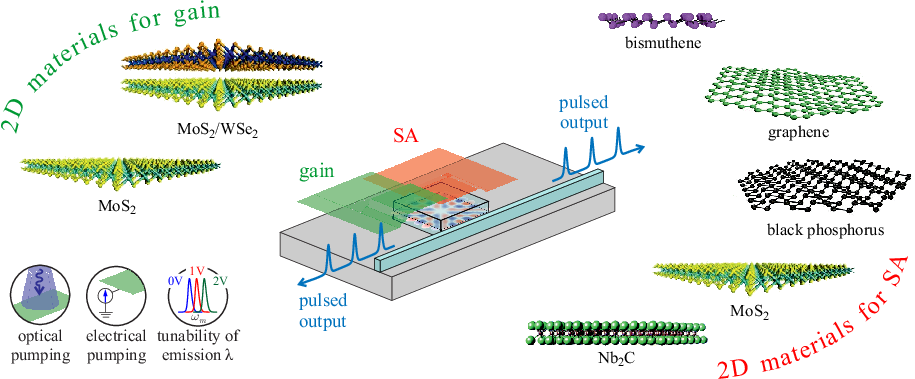}
        \caption{Generic nanophotonic resonant cavity with  gain (green sheet) and saturable absorption (red sheet) provided by different  2D materials. For appropriate pumping levels, the nanophotonic cavity can exhibit pulsed lasing, collected from an adjacent waveguide. The capability for either optical or electrical pumping and the tunability of the emission wavelength are important attributes of the considered platform.}
        \label{fig:Figure1}
    \end{figure}

    We first establish the numerical tools that will be used for studying the lasing response and performing a bifurcation/stability analysis of diverse nanophotonic $Q$-switched laser cavities incorporating 2D materials for gain and saturable absorption. It is critical that we use an accurate and efficient theoretical framework that allows to investigate and comparatively assess the performance of a broad range of 2D materials. A generic lasing cavity is depicted in Fig.~\ref{fig:Figure1}, along with the most prominent candidate 2D materials for gain and SA. 

\subsection{\label{subsec:FrameworkCMT}Coupled-mode theory equations for gain and saturable absorption}
    
    For the analysis of the nanophotonic resonator we will resort to temporal coupled-mode theory \cite{HausBook,Christopoulos2024Tut}, which can efficiently model gain and SA in nanophotonic cavities and systems \cite{Chua2011,Chua2014,Nousios2023,Nousios2024}. We choose CMT over traditional descriptions based on the density of the generated photons \cite{Siegman,KellerBook,Yamada1993}, since it allows to capture both the amplitude \emph{and} phase of the emitted light. Furthermore, CMT is necessary for the accurate incorporation of SA within a single, unified equation \cite{Ma2014,Rasmussen2017,Ataloglou2018,Nousios2022}. \blue{We elaborate on this choice by comparing our framework with another established approach for the $Q$-switching modeling in semiconductors, namely the \emph{Yamada model} \cite{Yamada1993,Dubbeldam1999}, in Appendix~\ref{app:AppC}.}
    
    We will present here only a normalized version of the CMT equations, which are more appropriate for numerical manipulations; the full set of equations, the rigorous calculation of the ODE coefficients from the actual physical system, the connection of the normalization constants to physical quantities and material properties, and the process of normalizing the involved variables are discussed in the Appendices; it is also thoroughly discussed in Refs.~\cite{Ataloglou2018,Nousios2023,Nousios2024}. 
    We will assume that gain is described by a simple two-level system. The framework can be readily extended to gain media with more levels \cite{Chua2014}, with the qualitative conclusions remaining largely unaffected. Ultimately, we can write the following set of ordinary differential equations (ODEs) \cite{Christopoulos2024Tut,Nousios2024}
    \begin{subequations}
        \begin{align}
            \fullFRAC{\tilde u}{t^\prime} &= (-j\delta - 1)\tilde u + g_1(\delta)\Delta\bar n\,\tilde u - r_\mathrm{SA}\left( 1 - \frac{\bar n_\mathrm{SA}}{2} \right)\tilde u, \label{eq:CMTdynSAAmpl} \\
            \fullFRAC{\Delta\bar n}{t^\prime} &= r_p - \frac{\Delta\bar n}{\tau_{21}^\prime} - g_2(\delta)\Delta\bar n|\tilde u|^2, \label{eq:CMTdynSADn} \\
            \fullFRAC{\bar n_\mathrm{SA}}{t^\prime} &= \frac{2}{\tau_\mathrm{SA}^\prime}\left[ 1 - \frac{\bar n_\mathrm{SA}}{2} \right] |\tilde u|^2 - \frac{\bar n_\mathrm{SA}}{\tau_\mathrm{SA}^\prime}, \label{eq:CMTdynnSA}
        \end{align}
        \label{eq:CMTdynSA}
    \end{subequations}
    where $\tilde u(t)$ is the slowly varying envelope of the normalized cavity amplitude $u(t) = \tilde u(t) \exp\{+j\omega_\mathrm{ref} t\}$, $\Delta\bar n(t)$ is the normalized spatially-averaged population inversion density in the gain medium, and $\bar n_\mathrm{SA}(t)$ is the normalized spatially-averaged free carrier density in the SA medium. The time variable, $t^\prime$, is also normalized with respect to the cavity lifetime; the prime symbol is used to denote normalization with respect to $\tau_\ell$. In the cavity amplitude equation~\eqref{eq:CMTdynSAAmpl}, $g_1(\delta)$ quantifies the strength of the gain mechanism \blue{and is, in general, a \textit{complex} quantity in detuned systems (see Appendices~\ref{app:AppA},~and~\ref{app:AppΒ})}, $r_\mathrm{SA}$ quantifies the level of unsaturated losses (alternatively the term normalized SA modulation depth may be used), and $\delta$ is a detuning parameter. Note that $\delta$ is \textit{not} the difference between the unperturbed (``cold'') resonance frequency of the cavity and the peak emission frequency of the gain medium, which is denoted with $\delta_\mathrm{out}$, \blue{and also implies an additional phase contribution in Eq.~\eqref{eq:CMTdynSAAmpl} through the complex $g_1(\delta)$ parameter}, see Appendices~\ref{app:AppA},~\ref{app:AppΒ},~\blue{and~\ref{app:AppC}}. In the population inversion density equation~\eqref{eq:CMTdynSADn}, $\tau_{21}^\prime$ is the lifetime of the carriers in the metastable level, $g_2$ is a gain coupling parameter, and $r_p$ is the normalized pumping intensity. Finally, in the SA carrier density equation~\eqref{eq:CMTdynnSA}, $\tau_\mathrm{SA}^\prime$ is the lifetime of the carriers in the SA medium. All the quoted parameters correspond to physical quantities through appropriate normalizations (see Appendix~\ref{app:AppA}).
    
    Importantly, with the introduction of detuning $\delta$, we can capture the general case where the peak emission frequency of the gain material, $\omega_m$, does not coincide with the resonance frequency of the cavity, $\omega_c$. As discussed in our previous work \cite{Nousios2023}, this case requires a careful treatment within the CMT framework. It turns out that $g_1$ and $g_2$ both depend on $\delta$ (intuitively anticipated since a detuning between $\omega_m$ and $\omega_c$ will limit the efficiency of light emission). In Appendix~\ref{app:AppΒ}, we discuss the optimum choice of $\delta$ for a given relation between $\omega_m$ and $\omega_c$, and how it is connected with $g_1(\delta)$ and $g_2(\delta)$. \blue{However, we shall note that the proposed methodology to choose $\delta$ is optimum only in terms of numerical efficiency and, generaly, any rational choice of the parameter leads to the same result \cite{Nousios2023}.}

    Equations~\eqref{eq:CMTdynSA} describe a quite general scenario. However, there are SA materials that respond almost instantaneously with respect to the cavity and population inversion lifetimes. This can simplify the set of Eqs.~\eqref{eq:CMTdynSA} after nullifying the derivative in Eq.~\eqref{eq:CMTdynnSA} and setting $\bar n_\mathrm{SA} = 2|\tilde u|^2/(1+|\tilde u|^2)$ in Eq.~\eqref{eq:CMTdynSAAmpl}. We can then write
    \begin{subequations}
        \begin{align}
            \fullFRAC{\tilde u}{t^\prime} &= (-j\delta - 1)\tilde u + g_1(\delta)\Delta\bar n\tilde u - \frac{r_\mathrm{SA}}{1+|\tilde u|^2}\tilde u, \label{eq:CMTinstaSAAmpl} \\
            \fullFRAC{\Delta\bar n}{t^\prime} &= r_p - \frac{\Delta\bar n}{\tau_{21}^\prime} - g_2(\delta)\Delta\bar n|\tilde u|^2. \label{eq:CMTinstaSADn}
        \end{align}
        \label{eq:CMTinstaSA}
    \end{subequations}
    The last term on the right-hand side of Eq.~\eqref{eq:CMTinstaSAAmpl} is now cast in the familiar form  $1/(1+x/x_\mathrm{sat})$.
    Equations~\eqref{eq:CMTinstaSA} represent a system with instantaneous SA, i.e., a system where $\tau_\mathrm{SA}^\prime \rightarrow 0$. The representation of gain in the two systems remains the same. In both Eqs.~\eqref{eq:CMTdynSA}~and~\eqref{eq:CMTinstaSA}, the sole driving term is $r_p$ (pumping intensity); it can correspond to either electrical or optical pumping.

    Note that Eqs.~\eqref{eq:CMTdynSA}~or~\eqref{eq:CMTinstaSA} should be in general complemented with an additional equation describing the out-coupling of light emitted by the cavity to the bus waveguide \cite{HausBook,Christopoulos2024Tut}. The specific form of this equation depends on the cavity type (standing/traveling wave)  and the coupling scheme (direct/side) \cite{Christopoulos2024Tut,Fan2002a}. However, in the absence of any other excitation but the pumping of the gain medium, the output power is directly proportional to the norm of the cavity amplitude squared and, thus, we can simply examine $|\tilde u|^2$ itself. 

\subsection{\label{subsec:FrameworkStab}Stability and bifurcation analysis of a laser cavity with saturable absorption}
    
    The simultaneous presence of gain and SA in a single cavity may result in an instability (specifically a limit cycle), leading to a stable pulsed output that is known as $Q$-switching \cite{Siegman,KellerBook,SalehBook}. The dynamic response of the system may be studied by solving the set of CMT Eqs.~\eqref{eq:CMTdynSA}~or~\eqref{eq:CMTinstaSA}, but such an approach is {obviously} slow and does not facilitate a qualitative understanding of the system under study. Alternatively, one can resort to other mathematical tools, such as {linear stability analysis and numerical continuation of solutions and bifurcations}, two approaches that inspect the system comprehensively and determine its dynamic behavior based on well established concepts such as the physical interpretation of its eigenvalues.
        
    Linear stability analysis can be used to evaluate the stability of a dynamical system near a fixed point $\mathbf{x}_0$, i.e., a CW solution of the system \cite{Rasmussen2017,Rasmussen2018,Kaminski2019}. Instead of solving a full system of nonlinear ODEs of the form $\mathbf{\dot x} = \mathbf{f}(\mathbf{x})$, one can linearize it using its Jacobian matrix $\mathbf{J}$ as $\mathbf{\dot x} = \mathbf{J}\mathbf{x}$. The Jacobian is calculated at the fixed point as $J_{ij}=\partial\{f_i(\mathbf{x})\}/\partial x_j|_{\mathbf{x}=\mathbf{x}_0}$. The stability of the linear system is determined by the eigenvalues of $\mathbf{J}$. It is then apparent that in order to determine the stability {of the CW solutions} of Eqs.~\eqref{eq:CMTdynSA}, one can find their Jacobian matrix and inspect its eigenvalues. In the simplest scenario, if at least one eigenvalue of $\mathbf{J}$ has a positive real part, then the respective fixed point is unstable. The type of instability depends on other factors, such as whether the eigenvalue is purely real or complex, the number of eigenvalues with positive real parts, or if they are simple or multiple. {Naturally, linear stability analysis and the examination of the eigenvalues can only provide information on the local dynamics around fixed points. For the treatment and characterization of periodic solutions ($Q$-switching) that involve the whole phase space of the system, we  employ a powerful software tool that executes a root-finding algorithm for continuation of solutions and bifurcations~\cite{Dhooge2003}. This  allows us to follow both stationary and periodic solutions, calculate their stability, and confirm the occurring bifurcations in a more effective 
    and mathematically rigorous manner compared to numerical integration.}

    \begin{figure}[!t]
        \centering
        \includegraphics[scale=1.0]{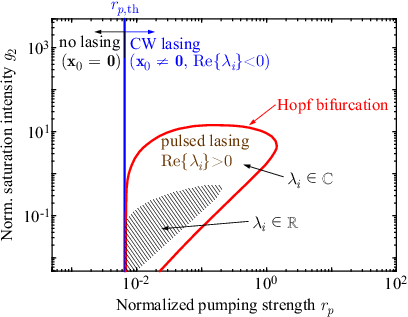}
        \caption{Bifurcation diagram of an indicative laser cavity with SA. Lasing is achieved above the lasing threshold $r_{p,\mathrm{th}} = 6.5\times10^{-3}$ (blue straight line). A Hopf bifurcation is found (red line), defining a limit cycles region and leading the system to a pulsed output ($Q$-switching). Results from linear stability analysis are also included in terms of the respective Jacobian's eigenvalues.}
        \label{fig:Figure2}
    \end{figure}
    
    An indicative bifurcation diagram  of a resonant cavity with gain and SA, \blue{calculated using Eqs.~\eqref{eq:CMTdynSA},} is depicted in Fig.~\ref{fig:Figure2}. We perform a {numerical continuation of the occurring bifurcation} in the $g_2$-$r_p$ parameter space and also include the findings of linear stability analysis, i.e., the nature of the eigenvalues of the respective Jacobian matrix. For the vertical axis, we have chosen the normalized parameter $g_2$ which directly appears in the CMT Eqs.~\eqref{eq:CMTdynSA}; it is proportional to $I_\mathrm{sat}$ (see Appendix~\ref{app:AppA}), which is the physically-relevant quantity (inherent property of the SA material). The (unitless) parameters used to obtain Fig.~\ref{fig:Figure2} are $\delta_\mathrm{out}=0$, $g_1=15.91$, $g_2=4.80\times10^{-11} I_\mathrm{sat}$, $\tau_{21}^\prime=110$, $r_\mathrm{SA}=9.09$, and $\tau_\mathrm{SA}^\prime=0.11$; therefore, the prefactor $4.80\times10^{-11}$ of $g_2$ has units of $\mathrm{m^2/W}$. They correspond to an indicative nanophotonic cavity with a TMD bilayer as the gain medium and graphene as the saturable absorber \cite{Nousios2024}. We will focus on alternative material options in section~\ref{sec:ExamplesBifurcation} but assume a similar configuration for the underlying cavity throughout.

    The results of Fig.~\ref{fig:Figure2} show that the parameter space is divided into three regions, defined by the blue and red lines. The blue line corresponds to the lasing threshold (either CW or pulsed), which is $r_{p,\mathrm{th}} = 6.5\times10^{-3}$. For $r_p<r_{p,\mathrm{th}}$, no lasing is possible since cavity loss dominates. In the terminology of dynamical systems, the only fixed point is the trivial solution $\mathbf{x}_0 = \mathbf{0}$. For $r_p > r_{p,\mathrm{th}}$ one obtains mostly CW lasing but there is a region defined by the red curve where a bifurcation occurs. This is a Hopf bifurcation, leading to a limit cycle for any point inside the red curve, rendering the output of the system pulsed ($Q$-switching). In linear stability analysis terms, the eigenvalues of the Jacobian outside(inside) the region defined by the red curve are complex conjugate with negative(positive) real parts, and the corresponding fixed point is a stable(unstable) focus. Exactly on the red line, the eigenvalue pair crosses the imaginary axis of the complex plane and a stable limit cycle is born via a Hopf bifurcation. This is a typical scenario for the generation of periodic solutions in nonlinear dynamical systems. 
    \blue{The observed Hopf instability is supercritical for the largest portion of the bifurcation line (i.e. the born limit cycle is stable), however there exists a narrow parameter range at low $r_p$ and $g_2$ values, where the Hopf is subcritical \cite{Guckenheimer2002,Strogatz2000}. The latter is only marginal and not addressed in the current study so we have chosen not to include it in the respective bifurcation diagrams.} 
    In addition, there is a smaller (hatched) area inside the region defined by the Hopf line, where the complex conjugate eigenvalues coalesce (exceptional point) and become real and positive, turning the unstable focus into an unstable node.  This change in the type of the fixed point does not affect the overall behavior in this region, which remains pulsed. However, more interesting phenomena are expected in more complex systems, e.g., in a system of two coupled cavities \cite{Ji2023}.

\section{\label{sec:ExamplesBifurcation} Bifurcation analysis to assess 2D material families for $Q$-switching}

    We will now use the tools of Sec.~\ref{sec:Framework} to assess the potential of various 2D materials that can provide gain and/or saturable absorption towards efficient nanophotonic $Q$-switched sources. 
    Regarding gain, most of the 2D material families have been investigated for light emission over the years. To date, the most promising family is transition metal dichalcogenides (TMDs), a family of 2D semiconductors that can emit at visible and near-infrared (NIR) wavelengths \cite{Wu2015,Ye2015}. Specifically, most of the available TMD monolayers (MoS$_2$, WS$_2$, WSe$_2$, InSe$_2$) have energy bandgaps corresponding to visible wavelengths and display two (low-energy) emission/absorption lines due to the formation of two excitonic transitions (A and B), accompanied with a ps lifetime in the metastable energy level and allowing for optical pumping and light emission in the visible. Even more interesting properties are found in the bilayers formed by two different TMDs (hetero-bilayers) that allow for an additional interlayer exciton transition between the conduction band of one and the valence band of the other, leading to light emission in the NIR  \cite{Ceballos2015,Miller2017,Karni2019,Paik2019}. MoS$_2$/WSe$_2$ bilayers, for example, have a peak emission frequency around 1100~nm (depending on the alignement angle between the two monolayers \cite{Paik2019,Lin2024}) and require optical pumping at 740~nm \cite{Karni2019}. Finally, TMD hetero-bilayers possess large carrier lifetimes in the metastable level (in the ns regime), which is quite important for efficient $Q$-switching \cite{Otupiri2020}, as will become evident in the following.
    
    Similarly, saturable absorption has been investigated with practically every available 2D material; they were found to exhibit noticeably different saturation intensities (spanning four or more orders of magnitude) and different lifetimes, ranging from the sub-ps (practically instantaneous) to the sub-ns regime. It is, thus, important to examine the appropriateness of a 2D material as a saturable absorber for $Q$-switching in the nanoscale. For each material we will use widely-accepted electromagnetic properties as reported in well-grounded literature resources. Note that in some cases the quality of the 2D material and the growth/transfer process can considerably impact its SA properties. TMD monolayers, for example, have been reported to exhibit saturation intensities between 10 and 100~MW/cm$^2$ and lifetimes from sub-ps to tens of ps \cite{Nie2014,Liu2014a,Zhang2016a,Cheng2016,Sun2017,Sun2019a}.
    
    The first and most well studied 2D material for SA is graphene, with a low saturation intensity even below 10~MW/cm$^2$ and a ps carrier lifetime \cite{Dawlaty2008,Bao2009,Xu2014}. Regarding other X-ene materials (X standing for materials in the IV and V groups of the periodic table), the most promising alternative appears to be the quantum-dot-enhanced bismuthene with very low saturation intensity (below 1~MW/cm$^2$) and a carrier lifetime around 10~ps \cite{Dong2020}. Other alternatives like antimonene, silicene, or germanene possess either similar or inferior properties to graphene \cite{Zhang2022,Song2017,Zhang2018b} but may provide other advantages like easier growth/transfer process or better compatibility with established PIC platforms. Black phosphorus (BP) is another interesting alternative; it has a similar carrier lifetime with graphene, an order of magnitude larger $I_\mathrm{sat}$, but it is relatively unstable when exposed to air \cite{Zhang2015b,Lu2015,Zhang2020b}. This can be improved however through encapsulation. 
    
    MX-enes and topological insulators (TIs) have also shown promise for SA. MX-enes consist of atomically thin layers of transition metal carbides or nitrides (M represents the transition metal and X stands for carbon or nitrogen), with the most well studied members of the family being Ti$_3$C$_2$T$_x$ and Nb$_2$C \cite{Li2019a,Gao2020}. The latter is a very good saturable absorber with almost instantaneous (sub-ps) response and low saturation intensity in the order of 1~MW/cm$^2$ \cite{Gao2020}. Regarding 2D TIs, thus far only Bi$_2$Te$_3$ has been shown to possess very low saturation intensity, even below 0.1~MW/cm$^2$ \cite{Wang2016a,Lin2016a}. This promising result has spurred the interest for TIs as saturable absorbers.

\subsection{\label{subsec:ExamplIsat}Saturation intensity}

    Initially, we will assume an indicative design for the underlying cavity and evaluate representative materials from each family to see whether they can be used to obtain $Q$-switching. The parameters shown in Table~\ref{tab:Table2} correspond to a photonic (dielectric) cavity with a resonance frequency that coincides with the peak emission frequency of the gain medium and possesses a very high intrinsic quality factor of $100\,000$. The lifetime of the cavity is therefore  dictated by its coupling strength with a bus waveguide, chosen to be relatively strong with an external $Q$-factor of $10\,000$ thus resulting in a loaded (total) quality factor of approximately $9\,000$. Such quality factors are achievable with integrated cavities \cite{Nousios2024}. Gain and SA sheets are placed in the vicinity of the cavity to interact with the resonant mode \cite{Nousios2023,Nousios2024}, see Fig.~\ref{fig:Figure1} for an indicative sketch. Only the SA material introduces ohmic loss; it is described by the parameter $r_\mathrm{SA} = 9.09$, which corresponds to a quality factor (for low intensities, i.e., in the absence of loss saturation) of around  $1\,000$. We should note here that, along with $I_\mathrm{sat}$ and $\tau_\mathrm{SA}$, SA materials are characterized by their modulation depth (MD), i.e., the ratio of saturable over non-saturable losses. This metric is captured by $r_\mathrm{SA}$ in our model and a large value of the said parameter corresponds to an optimum MD $\rightarrow 1$. We will see how the system behaves with respect to $r_\mathrm{SA}$ in section~\ref{subsec:ExamplrSA}. Here, we just stress that despite MD being an inherent property of the material, we can engineer $r_\mathrm{SA}$ (e.g., by appropriate placement of the SA medium) and, thus, maintaining a constant $r_\mathrm{SA} = 9.09$ for all SA materials is a rational choice.

    \begin{table}[!b]
        \centering
        \caption{Parameters used in CMT ODEs for Fig.~\ref{fig:Figure3}, which studies a nanophotonic cavity with SA and gain (TMD monolayer or hetero-bilayer).}
        \def\arraystretch{1.3}
        \setlength\tabcolsep{4.5pt}
        \begin{tabular}{lcccccc}
            \hline\hline
            ~ & $\delta_\mathrm{out}$ & $g_1$ & $g_2$ & $\tau_{21}^\prime$ & $r_\mathrm{SA}$ & $\tau_\mathrm{SA}^\prime$ \\
            \hline
            Monolayer & 0 & 15.91 & $4.80\times10^{-11} I_\mathrm{sat}$ & 1.1 & 9.09 & $^\mathrm{(a)}$ \\
            Bilayer   & 0 &  4.55 & $1.37\times10^{-11} I_\mathrm{sat}$ & 110 & 9.09 & $^\mathrm{(a)}$ \\
            \hline\hline
            \multicolumn{7}{l}{$^\mathrm{(a)}$\footnotesize{see the annotation in each panel of Fig. 3.}}
        \end{tabular}
        \label{tab:Table2}
    \end{table}
    
\begin{figure}[!t]
        \centering
        \includegraphics[scale=1.3]{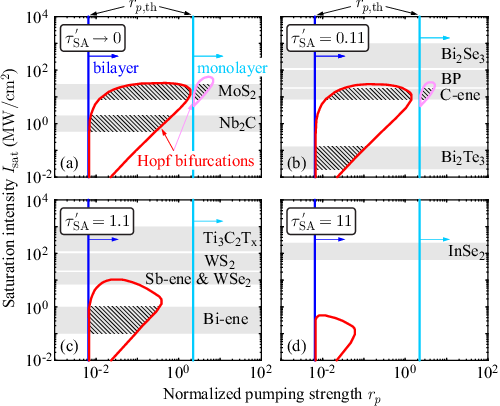}
        \caption{Bifurcation diagrams of a cavity with gain and SA in the $I_\mathrm{sat}$-$r_p$ parameter space, examining two different gain materials (TMD monolayer and bilayer) and assuming different carrier lifetimes for the saturable absorber: (a)~Instantaneous SA, (b)~$\tau_\mathrm{SA}=1~\mathrm{ps}$, (c)~$\tau_\mathrm{SA}=10~\mathrm{ps}$, and (d)~$\tau_\mathrm{SA}=100~\mathrm{ps}$. The CMT parameters are compiled in Table~\ref{tab:Table2}. A Hopf bifurcation and an instability region (limit cycles) that leads to  pulsed output ($Q$-switching) is present in all cases. In each subpanel, gray strips indicate the parameter range of a few 2D materials for SA. An overlap between a gray strip and an instability region (hatched areas) indicates that this 2D material can be utilized to achieve $Q$-switching.}
        \label{fig:Figure3}
    \end{figure}

    In Fig.~\ref{fig:Figure3}, we present the bifurcation analysis results for a lasing system with a TMD gain material (monolayer or hetero-bilayer). The CMT parameters that correspond to the physical system are compiled in Table~\ref{tab:Table2} \blue{and are used to populate either Eqs.~\eqref{eq:CMTdynSA}~or~\eqref{eq:CMTinstaSA}}. We examine  the $I_\mathrm{sat}$-$r_p$ parameter space and include four subpanels, one for an instantaneous saturable absorber [Fig.~\ref{fig:Figure3}(a), \blue{produced via Eqs.~\eqref{eq:CMTinstaSA}}] and three more for slower saturable absorbers with finite lifetimes, namely $\tau_\mathrm{SA} = \mathrm{\{1~ps,\,10~ps,\,100~ps\}}$ [Fig.~\ref{fig:Figure3}(b-d), \blue{produced via Eqs.~\eqref{eq:CMTdynSA}}]. For the vertical axis, we use the physically-meaningful material property $I_\mathrm{sat}$ (measured in $\mathrm{W/m^2}$), rather than the normalized (unitless) equivalent parameter  $g_2$ [see Appendix~\ref{app:AppA} and Table~\ref{tab:Table2}]. We consider two different options for the 2D gain material, one corresponding to a TMD bilayer (blue and red curves) and one to a TMD monolayer (cyan and pink curves). Qualitatively, in both cases the behavior remains similar with the generic system presented in Fig.~\ref{fig:Figure2}. However, in the bilayer case lasing is achieved for significantly lower pumping intensities. This is expected as the bilayer has a significantly larger metastable lifetime ($\tau_\mathrm{21,bi} = 1~\mathrm{ns}$ and $\tau_\mathrm{21,mono} = 10~\mathrm{ps}$) and it holds that $R_{p,\mathrm{th}} \propto 1/\tau_{21}$ \cite{Chua2011,Nousios2023}. Furthermore, $\tau_{21}$ also determines the ability of the system to achieve $Q$-switching; larger lifetimes (TMD bilayer) are favorable as the limit for $Q$-switching is inversely proportional to the population inversion lifetime of the gain medium \cite{KellerBook}. The opposite is true for the SA carriers lifetime. Faster saturable absorbers generally favor $Q$-switching, with pulsed output being achievable for a large range of $I_\mathrm{sat}$-$r_p$ sets. Systems with slower saturable absorbers generally require low $I_\mathrm{sat}$ values, which may be impractical or difficult to achieve. Moreover, saturable absorbers with large lifetimes recover more slowly, leading to smaller repetition rates \cite{KellerBook}. \blue{Finally, we note that the above discussion, but for semiconductor lasers, has led to similar observations \cite{Otupiri2020}: $Q$-switching is in general more favorable for systems with $\tau_\mathrm{SA}/\tau_{21} < 1$, but it can still be achieved even for $\tau_\mathrm{SA}/\tau_{21} > 1$ by fine-tuning the system parameters.}

    In each subpanel of Fig.~\ref{fig:Figure3}, we include with gray strips the range of experimentally specified $I_\mathrm{sat}$ values for different 2D saturable absorbers. An overlap between a gray strip and an instability region is denoted by the hatching and indicates that this 2D material can be utilized to achieve $Q$-switching for appropriate pumping levels. Evidently, graphene [C-ene label, Fig.~\ref{fig:Figure3}(b)] lies well within an instability region, either when a TMD monolayer or bilayer is used to provide gain. This is an expected result since in the literature there are several experimental demonstrations with graphene in fiber lasers \cite{Luo2010,Liu2011a,Popa2011}, although more rarely in PICs in the nanoscale \cite{Kovacevic2018,Hou2018}. Other 2D materials that seem to favor $Q$-switching in nanophotonics are Nb$_2$C and MoS$_2$ in Fig.~\ref{fig:Figure3}(a), Bi$_2$Te$_3$ in Fig.~\ref{fig:Figure3}(b), and bismuthene (labeled Bi-ene) in Fig.~\ref{fig:Figure3}(c). \blue{The trend is that fast saturable absorbers ($\tau_\mathrm{SA}/\tau_{21} \ll 1$) are generally favorable for demonstrating $Q$-switching in the nanoscale, since higher saturation intensities (corresponding to 2D material samples of lower quality) can be tolerated. In contrast, when $\tau_\mathrm{SA}/\tau_{21} > 0.1$ only high-quality samples can be utilized.}

    It should be kept in mind that the bifurcation maps of Fig.~\ref{fig:Figure3} correspond to a specific nanocavity and other configurations may favor other 2D materials. We will discuss such a case in the next section.  However, we have ensured realistic and indicative cavity parameters (specifically the overlap factors between the mode and gain/SA media) to allow reaching solid and general conclusions.

\subsection{\label{subsec:ExamplDet}Detuning of resonance and peak emission frequencies}

    In section~\ref{subsec:ExamplIsat} we have considered the simplest and most common case of $\omega_c = \omega_m$. In this section, we examine what happens when $\omega_c \neq \omega_m$, meaning that $\delta_\mathrm{out}$ (and thus $\delta$) will be nonzero. As discussed in Appendix~\ref{app:AppΒ}, when $\omega_c \neq \omega_m$, the lasing frequency ($\omega_L$) does not coincide with neither $\omega_m$ nor $\omega_c$; the actual lasing frequency can be estimated with good accuracy using Eq.~\eqref{eq:OmegaRef} of Appendix~\ref{app:AppΒ} [or Eq.~(25) of Ref.~\cite{Nousios2023} when only gain is present without SA].     Specifying $\omega_L$  is crucial for performing efficient bifurcation and stability analysis.

    \begin{figure}[!b]
        \centering
        \includegraphics[scale=1.05]{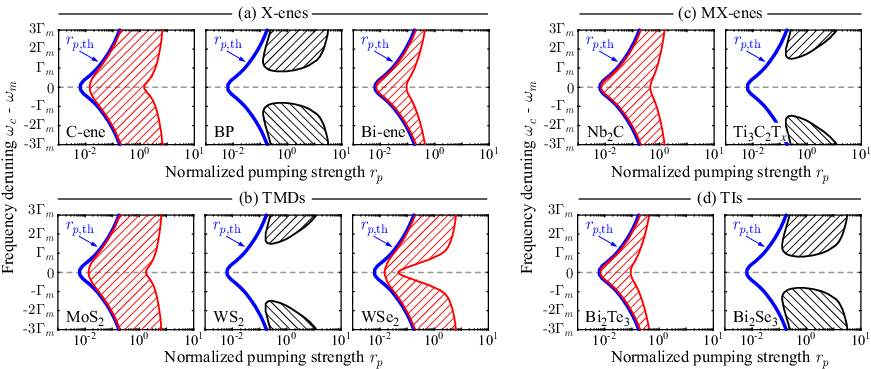}
        \caption{Bifurcation diagrams in the detuning-$r_p$ parameters space for a cavity with gain (TMD hetero-bilayer) and SA. Different material families for SA are examined: (a)~X-enes including graphene (C-ene), black phosphorus (BP) and bismuthene (Bi-ene). (b)~TMDs including MoS$_2$, WS$_2$, and WSe$_2$. (c)~MX-enes including  Nb$_2$C and Ti$_2$C$_2$T$_x$. (d)~TIs including Bi$_2$Te$_3$ and Bi$_2$Se$_3$. The parameters used are documented in Table~\ref{tab:Table3}. As detuning deviates from 0, (pulsed) lasing requires higher pumping intensities. Interestingly, detuning may allow for $Q$-switching operation with additional SA materials (see panels with black curves), at the expense of higher pumping intensities.}
        \label{fig:Figure4}
    \end{figure}

    We perform such an analysis and present the results in the detuning-$r_p$ parameter space in Fig.~\ref{fig:Figure4}. Several material families are  considered through some of their most promising members. The parameters to reproduce Fig.~\ref{fig:Figure4} are documented in  Table~\ref{tab:Table3}. All results are for a TMD bilayer gain medium. The vertical axis (detuning) is given in units of $\Gamma_m$, since $\omega_c$ must lie within the gain Lorentzian ($\Gamma_m$ is the full-width at half-maximum). The considered range ($\pm 3\Gamma_m$) corresponds to $\delta_\mathrm{out}\approx \pm 550$. Through Eqs.~\eqref{eq:OmegaRef}~and~\eqref{eq:g1g2Relation} of Appendix~\ref{app:AppΒ}, the required $\delta$, $g_1$, and $g_2$ parameters for the CMT ODEs~\eqref{eq:CMTdynSA} are retrieved.

    \begin{table}[!t]
        \centering
        \caption{List of parameters used in CMT ODEs for Fig.~\ref{fig:Figure4}, which studies the impact of detuning on the $Q$-switching regions for different SA materials. For their definitions, see Appendix~\ref{app:AppA}.}
        \def\arraystretch{1.3}
        \setlength\tabcolsep{4.5pt}
        \begin{tabular}{lcccccc}
            \hline\hline
            ~ & ~ & $\mathcal{A}$ & $\mathcal{B}$ & $\tau_{21}^\prime$ & $r_\mathrm{SA}$ & $\tau_\mathrm{SA}^\prime$ \\
            \hline
            X-enes & C-ene & 2\,892.6 & 871 & 110 & 9.09 & 0.11 \\
            ~ & BP & 2\,892.6 & 8\,712 & 110 & 9.09 & 0.11 \\
            ~ & Bi-ene & 2\,892.6 & 8.71 & 110 & 9.09 & 1.1 \\
            \hline
            TMDs & MoS$_2$ & 2\,892.6 & 871 & 110 & 9.09 & $\rightarrow 0$ \\
            ~ & WS$_2$ & 2\,892.6 & 8\,712 & 110 & 9.09 & 1.1 \\
            ~ & WSe$_2$ & 2\,892.6 & 871 & 110 & 9.09 & 1.1 \\
            \hline
            MX-enes & Nb$_2$C & 2\,892.6 & 87.1 & 110 & 9.09 & $\rightarrow 0$ \\
            ~ & Ti$_3$C$_2$T$_x$ & 2\,892.6 & 8\,712 & 110 & 9.09 & 1.1 \\
            \hline
            TIs & Bi$_2$Te$_3$ & 2\,892.6 & 8.71 & 110 & 9.09 & 0.11 \\
            ~ & Bi$_2$Se$_3$ & 2\,892.6 & 8\,712 & 110 & 9.09 & 0.11 \\
            \hline\hline
        \end{tabular}
        \label{tab:Table3}
    \end{table}

    The presented bifurcation curves are symmetric with respect to $\omega_c - \omega_m = 0$, a direct consequence of the symmetric Lorentzian lineshape of the gain medium. When $|\omega_c - \omega_m| \neq 0$, the  lasing threshold  shifts to higher pumping intensities (see blue curves in Fig.~\ref{fig:Figure4}). For most materials we find two Hopf bifurcation curves, setting the lower and the upper limit of the instability region (panels with red curves).    These materials can give pulsed output for any $\delta_\mathrm{out}$ value. Interestingly, we find instability regions for other materials, namely, BP, WS$_2$, Ti$_3$C$_2$T$_x$, and Bi$_2$Se$_3$ (panels with black curves), which are achieved for $|\delta_\mathrm{out}| > 0$. This shows that the suitable materials for $Q$-switching are not limited to those presented in Fig.~\ref{fig:Figure3}, but can be further expanded. By opting for $\delta_\mathrm{out} \neq 0$, one effectively tunes (decreases) the $g_1$ and $g_2$ parameters, which has been reported to modify the $Q$-switching appearance conditions \cite{KellerBook}. For example, to achieve $Q$-switching with  black phosphorus, one may set the detuning at approximately $2\Gamma_m$; however, at least one order-of-magnitude higher pumping intensity should be used compared to graphene for instance. Additionally, one should expect that the pulse-train characteristics (i.e., repetition rate, pulse width, peak amplitude, etc.) will be different; this will be discussed in section~\ref{sec:ExamplPulses}.

\begin{figure}[!t]
        \centering
        \includegraphics[scale=1.3]{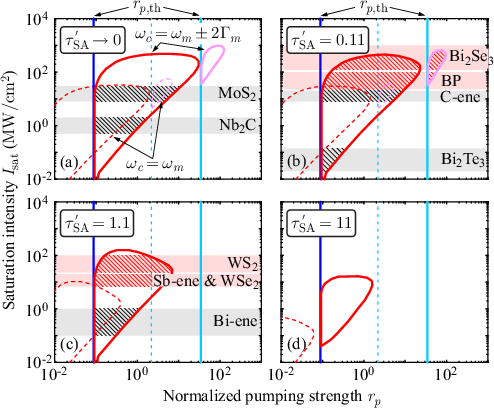}
        \caption{Bifurcation diagrams of a cavity with gain and SA in the $I_\mathrm{sat}$-$r_p$ parameter space, examining two different gain materials (TMD monolayer and bilayer, highlighted with blue/red and cyan/magenta curves, respectively), two different detuning values ($\omega_c = \omega_m$ and $\omega_c = \omega_m \pm 2\Gamma_m$), and assuming different carrier lifetimes for SA: (a)~Instantaneous SA, (b)~$\tau_\mathrm{SA}=1~\mathrm{ps}$, (c)~$\tau_\mathrm{SA}=10~\mathrm{ps}$, and (d)~$\tau_\mathrm{SA}=100~\mathrm{ps}$. With non-zero detuning one can achieve $Q$-switching for additional SA materials (compare dashed and solid lines defining the respective Hopf bifurcations or see the red-hatched regions). Black-hatched areas correspond to materials that give pulsed output regardless of detuning.}
        \label{fig:Figure5}
    \end{figure}

    The dependence of $Q$-switching regions with detuning can be further understood from the results in Fig.~\ref{fig:Figure5}, showing again bifurcation diagrams in the  $I_\mathrm{sat}$-$r_p$ parameter space, now including not only the optimum case $\omega_c = \omega_m$ (dashed lines) but also the case $\omega_c = \omega_m \pm 2\Gamma_0$ (solid lines). It is evident that detuning allows to achieve $Q$-switching with more SA material options (see the red-hatched areas in Fig.~\ref{fig:Figure5}), at the expense of higher pumping intensities. This is true even for the TMD monolayer option for gain where relatively larger $Q$-switching regimes emerge, but require even higher pumping. \blue{Therefore, tuning the distance between the cavity resonance and peak emission frequencies can further relax the requirements for fast and good-quality saturable absorbers. For example, through detuning we can achieve $Q$-switching with WS$_2$, a relatively slow and high-saturation-intensity saturable absorber ($\tau_\mathrm{SA}/\tau_{21} \sim 0.01$ and $I_\mathrm{sat} \sim 100~\mathrm{MW/cm^2}$). Importantly, 2D materials allow for tunability via, e.g., electrical gating; this way we can tune the emission spectrum and/or the resonance frequency of the underlying cavity.}
    
\subsection{\label{subsec:ExamplrSA}Modulation depth}

    Next, we select three promising SA materials and see how the normalized modulation depth, $r_\mathrm{SA}$, affects their ability to achieve $Q$-switching. We once again state that the $r_\mathrm{SA}$ parameter is analogous to the modulation depth of the SA medium, but can be engineered since it also depends on the overlap between the cavity mode and the material, i.e., on the degree of interaction between light and the SA medium. 
    Thus, one may balance the overall saturable and non-saturable losses of the system by, e.g., tailoring the position of the SA medium in the cavity.
    
\begin{figure}[!t]
        \centering
        \includegraphics[scale=1.05]{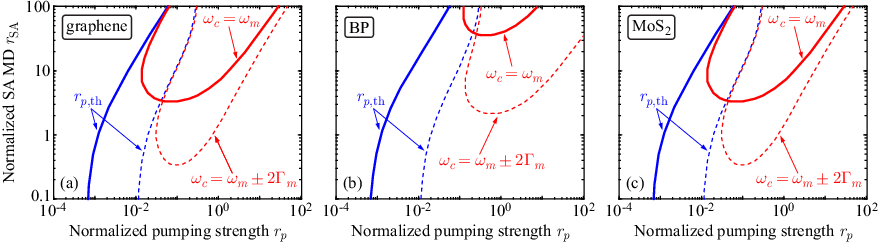}
        \caption{Bifurcation diagrams of a cavity with gain and SA in the $r_\mathrm{SA}$-$r_p$ parameter space, examining two different detuning values ($\omega_c = \omega_m$ and $\omega_c = \omega_m \pm 2\Gamma_m$) and three different SA materials: (a)~graphene, (b)~black phosphorus, and (c)~MoS$_2$. Pulsed output is realized inside the areas bounded by the red solid/dashed bifurcation curves. A higher level of saturable losses favors $Q$-switching but requires more pumping power. Pulsed output can be achieved even for $r_\mathrm{SA} < 1$ (higher non-saturable losses) by using a detuned cavity.}
        \label{fig:Figure6}
    \end{figure}

    We will focus on graphene, black phosphorus (BP), and molybdenum disulfide (MoS$_2$). The results of the bifurcation analysis are presented in the $r_\mathrm{SA}$-$r_p$ parameter space in Fig.~\ref{fig:Figure6}. The gain medium is a TMD bilayer. In each panel we study two cases for the cavity resonance frequency, the optimal case ($\omega_c = \omega_m$, solid lines) and a detuned one ($\omega_c = \omega_m \pm 2\Gamma_m$, dashed lines). Smaller values of $r_\mathrm{SA}$ correspond to lower saturable losses and thus allow lasing for lower pumping levels. However, small $r_\mathrm{SA}$ values also correspond to small modulation depths, rendering non-saturable and saturable losses comparable  and  hindering $Q$-switching.  In general, for achieving pulsed output the saturable part of the loss should be (significantly) larger than the non-saturable, otherwise the lasing is CW.  Note that with non-saturable loss we do not refer solely to the non-saturable ohmic loss of the SA material, but rather to any loss mechanism in the system that cannot be saturated. The configuration under study, for example, is dominated by coupling losses and this loss mechanism contributes significantly to the non-saturable part of the total quality factor. Finally, it is interesting to see that detuning can greatly expand the instability (pulsed) region, allowing for pulsed operation even when non-saturable losses are larger than the saturable ($r_\mathrm{SA}<1$), again at the expense of higher pumping intensities.

    \blue{The significance of the results presented in Fig.~\ref{fig:Figure6} should not be overlooked. Most studies thus far focus on large $r_\mathrm{SA}$ values, i.e., in systems where the saturable part of loss dominates. Indeed, such systems possess rich and interesting dynamics \cite{Dubbeldam1999,Otupiri2020}. However, here we show that there are cases where $Q$-switching can be achieved even when saturable losses are only a fraction of the system's total loss. In practice, this corresponds to cavities where the saturable absorber only weakly interacts with the lasing mode, thus providing additional flexibility in terms of cavity design, as well as sample quality and placement of the saturable absorber.}

\section{\label{sec:ExamplPulses}Pulse characteristics and time-domain evaluation of $Q$-switching}

\begin{figure}[!t]
        \centering
        \includegraphics[scale=1.05]{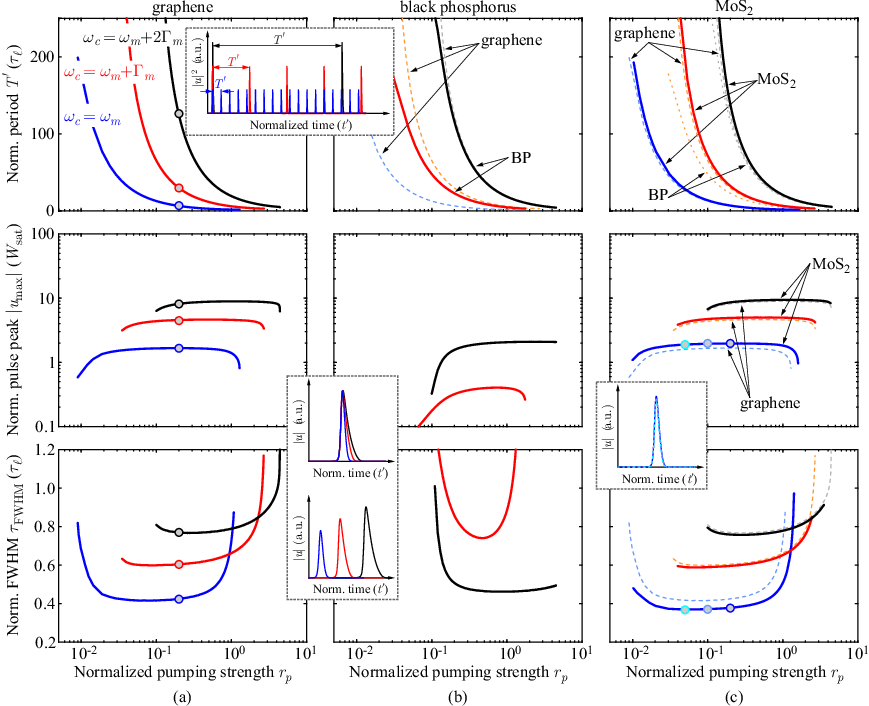}
        \caption{Characteristics of the emitted pulse-train (period, pulse peak, and FWHM, respectively) for three promising 2D saturable absorbers, assuming three different detuning values. (a) Graphene ($I_\mathrm{sat} = 10~\mathrm{MW/cm^2}$ and $\tau_\mathrm{SA}^\prime = 0.11$), (b) black phosphorus ($I_\mathrm{sat} = 100~\mathrm{MW/cm^2}$ and $\tau_\mathrm{SA}^\prime = 0.11$), and (c) MoS$_2$ ($I_\mathrm{sat} = 10~\mathrm{MW/cm^2}$ and $\tau_\mathrm{SA}^\prime \rightarrow 0$). Higher pumping intensities result in smaller periods (higher repetition rates). Cavity detuning also affects all the considered metrics, leading to sparser pulse-trains with longer and taller pulses. Additionally, it renders the pulses more asymmetric.  Both trends can be verified from the bottom left inset [note that $|u|$ is expressed in arbitrary units to facilitate comparison and each pulse corresponds to the respective dot in column (a)].}
        \label{fig:Figure7}
    \end{figure}

    Thus far, we have focused on the conditions that should be met to obtain a stable limit cycle. We will now turn our focus to the limit cycle itself and examine the characteristics of the emitted pulse-train. We will focus on the three SA materials that are mostly studied in the literature, namely, graphene, BP, and MoS$_2$. Gain will be provided by a TMD hetero-bilayer, e.g.,  MoS$_2$/WSe$_2$. In Fig.~\ref{fig:Figure7}, we present the obtained results (in normalized units) focusing on the period of the pulse-train  $T^\prime$ (inversely proportional to the repetition rate), the peak amplitude of each pulse $|\tilde u_\mathrm{max}|$, and its full-width at half-maximum (FWHM) $\tau_\mathrm{FWHM}$. Control over these metrics can be exerted by the pumping intensity $r_p$ (horizontal axis). The calculation of the period and peak amplitude has been conducted and confirmed via both integration of Eqs.~\eqref{eq:CMTdynSA} and numerical continuation of the respective limit cycles \cite{Dhooge2003}. Furthermore, we examine three different cases of detuning, namely $\omega_c = \omega_m$ (blue curves), $\omega_c = \omega_m + \Gamma_m$ (red curves), and $\omega_c = \omega_m + 2\Gamma_m$ (black curves), which correspond to $\delta_\mathrm{out} = 0$, $\delta_\mathrm{out} \approx 183$, and $\delta_\mathrm{out} \approx 366$, respectively. 

    Qualitatively, all configurations behave similarly and the same holds for any other material choices or combinations. Higher pumping intensities result in smaller periods or, equivalently, larger repetition rates. Single pulse characteristics, on the contrary, have a more complex behavior. The peak power initially increases before reaching an almost constant value for most of the $Q$-switching dynamic region (see the bottom right inset of Fig.~\ref{fig:Figure7}); it drops again as it reaches the upper cutoff limit of the limit cycle. Similarly, pulse duration first shortens rapidly before reaching an almost constant value (also see the bottom right inset) and then rises again close to the upper cutoff limit. Importantly, the cavity detuning can significantly affect the examined metrics, leading to larger periods (lower repetition rates), but with longer and taller pulses, i.e., carrying more energy.  This behavior can be easily understood since a lower repetition rate allows for more energy to accumulate in the cavity before it is ultimately released in the form of a pulse. Finally, as is evident from the bottom left insets, more energy accumulation inside the cavity before its ultimate release leads to asymmetric pulses with a steep leading edge (note that all insets of Fig.~\ref{fig:Figure7} are expressed in arbitrary units to facilitate comparison). This is a well-known behavior in $Q$-switched lasers \cite{Siegman,KellerBook}. \blue{One may observe a similar asymmetry in terms of modulation depth; in a system that saturable losses dominate ($r_\mathrm{SA}\rightarrow\infty$), the emerging pulses are strongly asymmetric \cite{Erneux2000}.} 
    
    Quantitatively, the choice of saturable absorber significantly impacts the metrics of the produced pulse-train. Graphene and BP, for example, have similar lifetimes but an order of magnitude different saturation intensities. The higher $I_\mathrm{sat}$ of BP leads almost always to pulses with lower peak power and shorter duration, but similar periods. For $\delta_\mathrm{out} = 0$ we do not get pulsed output and thus the respective blue curve is missing. On the other hand, MoS$_2$ is a faster saturable absorber than both graphene and BP, a fact that leads to pulse-trains with slightly larger periods (slightly lower repetition rate), and single pulses with slightly higher peak powers and slightly smaller widths, a direct consequence of the faster response of the system; cf. solid and dashed lines in column (c). Generally speaking, the carrier lifetime of the saturable absorber does not impact the shape of the pulse, which is typically symmetric and of sech$^2$ shape \cite{KellerBook,Erneux2000}. \blue{On the other hand, the relation between saturable and non-saturable losses affects the shape of the pulse; asymmetric shapes arise when saturable loss dominate, i.e., for $r_\mathrm{SA} \gg 1$, whereas in systems with better balance between saturable and non-saturable losses ($r_\mathrm{SA}\sim 1$-$10$), output pulses tend to be more symmetric (again of sech$^2$ shape) \cite{KellerBook,Erneux2000}.} As $\delta_\mathrm{out}$ increases, a similar asymmetry emerges, this time due to the limitation of the provided gain for the same level of pumping because of  suboptimal coupling with the resonance cavity. As an example, see the bottom left inset in Fig.~\ref{fig:Figure7} with the respective pulses for graphene.

    We should stress that the exact characteristics of the pulse-train depend on the properties of the gain and the SA media, as well as on the design of the cavity. 
    \blue{However, there are some universal trends (some of which not directly deducted from Fig.~\ref{fig:Figure7} but validated via additional simulations not shown) summarized as follows: A fast saturable absorber ($\tau_\mathrm{SA} < 1~\mathrm{ps}$) generally results in shorter pulses with higher peak power but smaller repetition rate. Similarly, high-quality saturable absorbers with low saturation intensities ($I_\mathrm{sat} < 10~\mathrm{MW/cm^2}$) are associated with significantly higher peak power and typically longer duration, i.e., with pulses carrying more energy. Finally, the level of saturable (compared to the non-saturable) losses also affects the pulse-train characteristics, with large $r_\mathrm{SA}$ values (domination of saturable losses) resulting in shorter pulses with higher peak power but significantly lower repetition rate.}
    Furthermore, one can achieve a range of pulse widths, amplitudes, and repetition rates  by simply selecting an appropriate pumping level \blue{or even detuning between the cavity resonance and the peak emission frequencies.} In the next, concluding section, we will discuss the range of metrics possible with the examined platform and assess its perspective for practical nanophotonic lasers. 

\section{\label{sec:Concl}Discussion and Conclusions}

    In this section, we will compare the indicative results of Fig.~\ref{fig:Figure7} with those of other configurations for $Q$-switching. The proposed platform of an integrated nanophotonic cavity enhanced with 2D materials for gain and SA can offer repetition rates ranging from a few hundred MHz up to 50~GHz, by simply tuning the pump intensity. Short pulses with ps duration (as low as 3~ps for the MoS$_2$ or 3.5~ps for graphene) can be generated for most of the aforementioned repetition rates. Finally, the energy of the output pulse depends on the coupling strength between the cavity and the bus waveguide. For a relatively strongly coupled system (coupling loss 10 times higher than linear intrinsic loss), pulses with energy from a few to one hundred fJ can be generated. Higher pulse energies correspond to asymmetric pulses achieved for large detunings. The peak power can be in the order of few mW. 
    
    We should stress how different are the $Q$-switching characteristics of our 2D-material-enhanced integrated lasers, compared to those of 2D-material-based fiber lasers, which exploit $Q$-switching as well but are are characterized by significantly longer roundtrip times \cite{Tiu2019}. Fiber laser implementations are characterized by much slower repetition rates (tens of kHz up to few MHz) and longer pulses (hundreds of ns up to $\mu$s) which, however, carry significantly higher optical energy (tens of nJ or higher) and may reach peak powers $>1$~W \cite{Tiu2019,Zhang2020}. The connection discussed earlier between repetition rate, pulse width, and pulse energy (or peak power) is maintained (smaller repetition rates lead to longer and taller pulses due to greater energy accumulation).
    
    2D-material-based saturable absorbers have been also used in external cavity lasers, where light is provided in these cavities by external bulk laser crystals (neodymium, erbium, or ytterbium) \cite{Zhang2019} or external solid-state lasers \cite{Hao2020}.  Monolithically-integrated waveguide lasers where the 2D SA material is placed in the vicinity of the waveguide and the gain is provided by doping the waveguide with Nd, Er, or Yt ions have been also examined \cite{Li2020b}. In the aforementioned cases, the cavities are typically of mm size or longer; the resulting pulses are of tens of ns duration, since the pulse FWHM scales with the cavity roundtrip \cite{KellerBook,Sphler1999}. There have been realizations of smaller cavities utilizing, for example, semiconductor saturable absorption mirrors (SESAMs), and in such systems pulses with tens of ps duration have been predicted and measured \cite{Sphler1999,Butler2012,Yu2017b}. With regards to pulse-train period, slow repetition rates ranging from kHz to MHz are achieved, as is mostly dictated by the ms population inversion lifetimes of the gain media (rare earth ions or III-V semiconductors). Specifically, the  period of the pulse-stream is proportional to $\tau_{21}$ \cite{KellerBook,Sphler1999}.
    
    It therefore becomes evident why our considered system consisting of $\mu$m-scale cavities and gain media with ns population inversion lifetimes manages to achieve attractive performance metrics (GHz repetition rate \emph{and} ps duration), filling an important gap in the parameter space of pulsed laser sources \cite{Charlet2011,Yu2017b,Shtyrkova2019}. 2D materials contribute to this goal in multiple ways: (i)~As gain media, apart from the favorable lifetime, they offer the capability for either optical (photoluminesence) or electrical  pumping (electroluminescence) \cite{Chen2024arXiv}. Furthermore, most of the considered 2D materials have electrically tunable properties, allowing for additional freedom in achieving detuned scenarios through the electrical control of the  emission frequency. (ii)~As saturable absorbers, they offer broadband response, in contrast with rather narrowband and bulkier SESAMs that have been previously used. 
    Moreover, the loss saturation level or the ratio of saturable versus non-saturable losses (modulation depth) can also be electrically tuned, allowing for further control in the $Q$-switching process.
    
    In conclusion, addressing the question posed in the title of the paper we found that enhancing compact passive nanophotonic cavities with 2D gain and SA media which are technologically compatible with  established PIC platforms (silicon, silicon nitride, III-V semiconductors, etc.) allows for achieving ultracompact integrated lasers with attractive pulse-train metrics that have not been attained with other approaches and also enables electrical tunability of the emission characteristics. 
    \blue{We acknowledge that achieving the calculated metrics in actual samples depends on the fabrication quality. Thus, experimental challenges such as the preparation of uniform and high-quality 2D material samples and the accurate control of their properties and placement in the cavity should not be overlooked. However, with the progress of nanofabrication, 2D-material growth/deposition and PIC technology in general, we still foresee a strong potential for $Q$-switching with 2D materials in the nanoscale. This belief is driven not only by the promising results presented in this work showing that indeed $Q$-switching is achievable with a multitude of 2D materials and in a wide range of parameters, but also by the flourishing studies, both theoretical and experimental, involving 2D materials that are found in the literature.}

\section*{Acknowledgements}
This research work was supported by the Hellenic Foundation for Research and Innovation (H.F.R.I.) under the ``2nd Call for H.F.R.I. Research Projects to support Post-doctoral Researchers'' (Project Number: 916, PHOTOSURF).

\appendix

\section{\label{app:AppA}CMT equations and normalization}
\renewcommand{\theequation}{A\arabic{equation}}
\setcounter{equation}{0}

    In this section, we will present the complete CMT expressions for gain and saturable absorption, their connections with the actual physical quantities that describe the laser cavity, and the normalization approach that leads to Eqs.~(1) of section~2 in the main text. Starting for the CMT amplitude differential equation, we have \cite{Ataloglou2018,Nousios2023,Christopoulos2024Tut}
    \begin{align}
        \fullFRAC{\tilde a(t)}{t} = -&j(\omega_\mathrm{ref}-\omega_c) \tilde a(t) - \left(\frac{1}{\tau_i}+\frac{1}{\tau_e}\right) \tilde a(t) + \nonumber \\
        +&\xi_1\frac{j\omega_\mathrm{ref}\sigma_m}{\omega_m^2-\omega_\mathrm{ref}^2+j\omega_\mathrm{ref}\Gamma_m}\Delta\bar N(t) \tilde a(t) - \frac{1}{\tau_\mathrm{SA,0}}\left[1-\frac{\bar N_\mathrm{SA}(t)}{2N_\mathrm{sat}}\right]\tilde a(t).
        \label{eq:CMTinitial}
    \end{align}
    The first two terms on the right-hand side represent the linear response of the cavity (resonance and losses, respectively), the third term the gain mechanism, and the fourth the saturable absorption process. Specifically, in Eq.~\eqref{eq:CMTinitial} $a(t) = \tilde a(t)\exp\{+j\omega_\mathrm{ref}t\}$ is the amplitude of the photonic mode, normalized so that it equals the stored energy in the cavity, i.e., $|a|^2 \equiv W_\mathrm{res}$ and $\tilde a(t)$ is a slowly varying envelope. $\Delta\bar N(t)$ is the spatially-averaged population inversion density in the gain medium, and $\bar N_\mathrm{SA}(t)$ is the spatially-averaged free carrier density in the SA medium. Equation~\eqref{eq:CMTinitial} does not describe the most general case of a gain medium (it omits, for example, the description via a polarization field, see Refs.~\cite{Nousios2023,Chua2011}) and also does not include other nonlinearities, like the Kerr effect. It is written in its simplest possible form to accurately describe the considered phenomena.
    
    The nanophotonic resonant cavity is characterized by a ``cold'' resonance frequency $\omega_c$ and a loaded lifetime $\tau_\ell = (1/\tau_i+1/\tau_e)^{-1}$, with $\tau_i$ and $\tau_e$ being the intrinsic and external lifetimes, associated with losses (radiation, ohmic) and coupling, respectively. Furthermore, each lifetime is connected with the respective quality factor through $Q=\omega_c\tau/2$ \cite{Christopoulos2019}. The gain medium exhibits a peak emission frequency $\omega_m$ and a gain linewidth $\Gamma_m$, described by a homogeneously broadened Lorentzian oscillator model. $\xi_1$ is an overlap factor between the gain medium and the photonic mode \cite{Nousios2023} and $\sigma_m$ is a coupling parameter that characterizes the gain medium and generally depends on the emission wavelength and the spontaneous emission rate \cite{Siegman}. Note that $\xi_1$ can be rigorously calculated if the resonant mode is known (e.g., after modal simulations \cite{Christopoulos2024Tut}) via appropriate overlap integrals; analytical expressions for $\xi_1$ can be found in Refs.~\cite{Chua2011,Nousios2023}.
    Finally, the SA material is characterized by the carrier saturation intensity $N_\mathrm{sat}$, being a measure of the free carrier density needed to suppress optical absorption to 50\%, and $\tau_\mathrm{SA,0}$ is the ohmic loss lifetime of the SA medium before loss saturation, i.e., for low input intensities. If the SA medium has also a non-saturable linear loss mechanism, its contribution should be included in the intrinsic lifetime $\tau_i$.
    
    There is one more parameter that was deliberately not discussed so far: the reference frequency $\omega_\mathrm{ref}$, which we have used as the time-harmonic optical frequency. In Ref.~\cite{Nousios2023}, its necessity is discussed in depth but we briefly state here that, when the cavity resonance $\omega_c$ does not coincide with the peak emission frequency $\omega_m$, the emerging lasing frequency $\omega_L$ does not coincide with neither   $\omega_c$ nor   $\omega_m$. In Ref.~\cite{Nousios2023} a method to estimate $\omega_L$ with quite good accuracy is presented but for a system with gain only; in section~\ref{app:AppΒ} we discuss how to estimate $\omega_L$ in a system with gain \emph{and} SA. In Eq.~\eqref{eq:CMTinitial}, $\omega_\mathrm{ref}$ is used instead of $\omega_L$ to capture the most general case and it can be chosen either arbitrarily (but reasonably close to $\omega_m$ and/or $\omega_c$) or one can use the estimation for $\omega_L$ as $\omega_\mathrm{ref}$. The latter is preferable and it has been shown to result in a more efficient solution of the CMT coupled ODEs \cite{Nousios2023}. When $\omega_c = \omega_m$, then $\omega_L = \omega_m$ as well, and any option for $\omega_\mathrm{ref}$ different that $\omega_c$ is redundant.
    
    The population inversion in the gain medium, $\Delta N(t,\mathbf{r})$, is governed by a partial differential equation (PDE) for 2-level media or emerges though a system of coupled PDEs in media with more energy levels \cite{Siegman}. For simplicity, we will examine the simplest  scenario of a 2-level gain medium and present the differential equation that describes not $\Delta N(t,\mathbf{r})$ but its spatially averaged equivalent quantity $\Delta\bar N(t)$ \cite{Chua2011,Nousios2023},
    \begin{equation}
        \fullFRAC{\Delta\bar N(t)}{t} = R_p - \frac{\Delta\bar N(t)}{\tau_{21}} - \frac{\xi_2}{\hbar\omega_m}\frac{1}{2}\mathrm{Re}\left\{\frac{j\omega_\mathrm{ref}\sigma_m}{\omega_m^2-\omega_\mathrm{ref}^2+j\omega_\mathrm{ref}\Gamma_m}\right\}\Delta\bar N(t) |\tilde a(t)|^2.
        \label{eq:DeltaNinitial}
    \end{equation}
    The first terms represents the pumping intensity, the second corresponds to non-radiative recombination, and the third describes stimulated emission following a homogeneously broadened Lorentzian oscillator model.
    Specifically, in Eq.~\eqref{eq:DeltaNinitial}, $\tau_{21}$ is the carrier lifetime in the metastable level (in the same order as the spontaneous emission lifetime) and of course should be large enough to allow for population inversion conditions. Additionally, $\xi_2$ is another overlap factor between the photonic mode and the gain material, different from $\xi_1$ and consistent with the chosen form of the spatial averaging for $\Delta\bar N$ \cite{Nousios2023,Nousios2024}, and $\hbar\omega_m$ is the energy of a single emitted photon. As with $\xi_1$, analytical expressions for $\xi_2$ can be found in Refs.~\cite{Chua2011,Nousios2023}. Finally, $R_p$ is the pumping intensity, deliberately kept in this general form which can be associated  with either electrical or optical pumping. We also denote with $N_\mathrm{tot}$ the averaged total number of available carriers (i.e., the total carrier density) in the gain medium. 
    
    The free carrier density in the SA medium $N_\mathrm{SA}(t,\mathbf{r})$ is similarly governed by another PDE. In the same spirit, we use the respective equation for the spatially averaged quantity $\bar N_\mathrm{SA}(t)$, which is \cite{Nousios2022,Christopoulos2024Tut}
    \begin{equation}
        \fullFRAC{\bar N_\mathrm{SA}(t)}{t} = \xi_3\left[1-\frac{\bar N_\mathrm{SA}(t)}{2N_\mathrm{sat}}\right]|\tilde a(t)|^2 -\frac{\bar N_\mathrm{SA}(t)}{\tau_\mathrm{SA}}.
        \label{eq:NSAinitial}
    \end{equation}
    The first terms represents optical absorption and its saturation process as free carriers accumulate in the conduction band of the SA medium, whereas the second describes carrier relaxation back in the valence band.
    Specifically, in Eq.~\eqref{eq:NSAinitial}, $\tau_\mathrm{SA}$ is the carrier lifetime of the SA medium and $\xi_3$ is a third overlap factor, now between the photonic mode and the SA material that quantifies the strength of  optical absorption  \cite{Ataloglou2018,Christopoulos2024Tut}.
    
    A comparison between Eqs.~\eqref{eq:CMTinitial}-\eqref{eq:NSAinitial} and Eqs.~(1) of the main text reveals that one can reach the latter after a proper normalization of the former. To do so, we define the normalized quantities $u$, $\Delta\bar n$, and $\bar n_\mathrm{SA}$ for the cavity amplitude, population inversion density, and SA free carrier density through
    \begin{subequations}
        \begin{align}
            u(t) &= \frac{a(t)}{\sqrt{W_\mathrm{sat}}}, \\
            \Delta\bar n(t) &= \frac{\Delta\bar N(t)}{N_\mathrm{tot}}, \\
            \bar n_\mathrm{SA}(t) &= \frac{\bar N_\mathrm{SA}(t)}{N_\mathrm{sat}},
        \end{align}
        \label{eq:NormalizationDefinitions}
    \end{subequations}
    respectively. It should be noted that the cavity amplitude is normalized with respect to the saturation energy $W_\mathrm{sat}$, connected with the carrier saturation density $N_\mathrm{sat}$ as $W_\mathrm{sat} = 2N_\mathrm{sat}/(\xi_3\tau_\mathrm{SA})$. More importantly, carrier saturation density $N_\mathrm{sat}$ is also connected with the saturation intensity $I_\mathrm{sat}$ through $I_\mathrm{sat} = (2\hbar\omega/\sigma_0\eta_0\tau_\mathrm{SA}) N_\mathrm{sat}$; $\eta_0 = 120\pi~\Omega$ is the free space impedance and $\sigma_0$ is the unsaturated conductivity of the SA material \cite{Chatzidimitriou2020}. It should be emphatically stated that $I_\mathrm{sat}$ (in $\mathrm{W/m^2}$) is the parameter that can be measured experimentally, thus it is widely used in the literature to characterize the saturable-absorption capabilities of a material. Finally, the time variable is also normalized with respect to the loaded cavity lifetime, i.e., $t^\prime = t/\tau_\ell$.
    
    It is now straightforward to reach Eqs.~(1) of the main text after the introduction of Eqs.~\eqref{eq:NormalizationDefinitions} into \eqref{eq:CMTinitial}-\eqref{eq:NSAinitial}. The resulting equations are repeated here for easier reference
    \begin{subequations}
        \begin{align}
            \fullFRAC{\tilde u}{t^\prime} &= (-j\delta - 1)\tilde u + g_1\Delta\bar n\tilde u - r_\mathrm{SA}\left( 1 - \frac{\bar n_\mathrm{SA}}{2} \right)\tilde u, \label{eq:CMTdynSAAmplApp} \\
            \fullFRAC{\Delta\bar n}{t^\prime} &= r_p - \frac{\Delta\bar n}{\tau_{21}^\prime} - g_2\Delta\bar n|\tilde u|^2,\\
            \fullFRAC{\bar n_\mathrm{SA}}{t^\prime} &= \frac{2}{\tau_\mathrm{SA}^\prime}\left( 1 - \frac{\bar n_\mathrm{SA}}{2} \right) |\tilde u|^2 - \frac{\bar n_\mathrm{SA}}{\tau_\mathrm{SA}^\prime}. 
        \end{align}
        \label{eq:CMTdynSAApp}
    \end{subequations}
    The normalized parameters are given by the following expressions:
    \begin{subequations}
        \begin{align}
            \delta &= (\omega_\mathrm{ref}-\omega_c)\tau_\ell, \label{eq:NormalQuantDet} \\
            g_1 &= \xi_1\frac{j\omega_\mathrm{ref}\sigma_m}{\omega_m^2-\omega_\mathrm{ref}^2+j\omega_\mathrm{ref}\Gamma_m} \tau_\ell N_\mathrm{tot}, \label{eq:NormalQuantg1} \\
            g_2 &= \frac{\xi_2}{\hbar\omega_m}\frac{1}{2}\mathrm{Re}\left\{\frac{j\omega_\mathrm{ref}\sigma_m}{\omega_m^2-\omega_\mathrm{ref}^2+j\omega_\mathrm{ref}\Gamma_m}\right\} \tau_\ell W_\mathrm{sat}, \\
            r_p &= \frac{R_p\tau_\ell}{N_\mathrm{tot}}, \\
            \tau_{21}^\prime &= \frac{\tau_{21}}{\tau_\ell}, \\
            r_\mathrm{SA} &= \frac{\tau_\ell}{\tau_\mathrm{SA,0}}, \\
            \tau_\mathrm{SA}^\prime &= \frac{\tau_\mathrm{SA}}{\tau_\ell}.
        \end{align}
        \label{eq:NormalQuant}
    \end{subequations}
    All quantities are purely real except for $g_1$ which is rendered complex when $\omega_\mathrm{ref}\neq\omega_m$. Furthermore, $\omega_\mathrm{ref}$ appears in both $g_1$ and $g_2$ and, thus, these parameters depend on our choice of $\delta$.
    Consequently, it is convenient to define a more physically meaningful detuning parameter $\delta_\mathrm{out} = (\omega_c-\omega_m)\tau_\ell$, which will be useful for the calculations in the following section~\ref{app:AppΒ}. Obviously, $\delta_\mathrm{out} = -\delta + (\omega_\mathrm{ref}-\omega_m)\tau_\ell$ and their relation depends on the exact choice of $\omega_\mathrm{ref}$, also affecting the values of $g_1$ and $g_2$.

\section{\label{app:AppΒ}Accurate description of detuning in systems with gain and SA}
\renewcommand{\theequation}{B\arabic{equation}}
\setcounter{equation}{0}

    As discussed in section~\ref{app:AppA}, one can readily handle Eqs.~\eqref{eq:CMTdynSAApp} by making a reasonable choice for $\omega_\mathrm{ref}$ (or, effectively, for $\delta$) and study the cavity dynamics. However, there is always an optimum choice for $\omega_\mathrm{ref}$ and $\delta$ that results in a more efficient way to handle the emerging system of ODEs and this choice is $\omega_\mathrm{ref} \equiv \omega_L$.
    
    As shown in Ref.~\cite{Nousios2023}, when $\omega_c \neq \omega_m$ one can estimate the lasing frequency $\omega_L$ and appoint that value to the cavity amplitude CMT ODE as $\omega_\mathrm{ref}$. Although not strictly necessary, this option simplifies the numerical solution since when $\omega_\mathrm{ref} \neq \omega_L$, both real and imaginary parts of $\tilde u$ oscillate at a frequency $|\omega_L-\omega_\mathrm{ref}|$ but with a 90$^\mathrm{o}$ phase difference. Due to this constant phase difference, the oscillations are not transferred to $|\tilde u|$ and, strictly speaking, any choice of $\omega_\mathrm{ref}$ correctly captures light emission. Obviously, the option  $\omega_\mathrm{ref} = \omega_L$ always eliminates these oscillations and simplifies the numerical solution by allowing for larger steps in the Runge-Kutta algorithm.
    
    But how can we estimate $\omega_L$? Intuitively, one may expect that $\omega_L$ equals $\omega_c$ or $\omega_m$, which is true only in the trivial but quite common case of $\omega_c = \omega_m$. In this specific case, $\delta_\mathrm{out} = 0$ and $\delta = 0$ is the optimum choice.  In the general case of $\delta_\mathrm{out} \neq 0$, however, one may estimate $\delta$ by solving Eq.~\eqref{eq:CMTdynSAAmplApp} for $\Delta\bar n$ in CW, obtaining
    \begin{equation}
        \Delta\bar n = \frac{\left[1+r_\mathrm{SA}\left(1-\displaystyle\frac{\bar n_\mathrm{SA}}{2}\right)\right]+j\delta}{g_1}.
        \label{eq:DeltaNCW}
    \end{equation}
    $\Delta\bar n$ is a strictly real quantity and imposing this restriction on Eq.~\eqref{eq:DeltaNCW}, one finds
    \begin{equation}
        \delta~\mathrm{Re}\{g_1\} = \mathrm{Im}\{g_1\}\left[1+r_\mathrm{SA}\left(1-\frac{\bar n_\mathrm{SA}}{2}\right)\right],
        \label{eq:g1Relation}
    \end{equation}
    a condition that should always be met. This condition ensures that $\Delta\bar n$ is always real and simultaneously eliminates the aforementioned oscillations of $\tilde u$. Now, since both $\delta$ and $g_1$ depend on $\omega_\mathrm{ref}$, Eq.~\eqref{eq:g1Relation} can be used to estimate $\omega_\mathrm{ref}$ through
    \begin{equation}
        \Omega_\mathrm{ref} = \frac{G_m(\delta_\mathrm{out}+\Omega_m) + \sqrt{G_m^2(\delta_\mathrm{out}+\Omega_m)^2+4\Omega_m^2\left[1+G_m+r_\mathrm{SA}\left(1-\displaystyle\frac{\bar n_\mathrm{SA}}{2}\right)\right]\left[1+r_\mathrm{SA}\left(1-\displaystyle\frac{\bar n_\mathrm{SA}}{2}\right)\right]}}{2\left[1+G_m+r_\mathrm{SA}\left(1-\displaystyle\frac{\bar n_\mathrm{SA}}{2}\right)\right]}.
        \label{eq:OmegaRef}
    \end{equation}
    We use the notation $\Omega = \omega\tau_\ell$ and $G_m = \Gamma_m\tau_\ell$ to denote the respective normalized (with respect to the cavity lifetime $\tau_\ell$) quantities. In Eq.~\eqref{eq:OmegaRef}, we have also used the fact that $\Omega_c = \delta_\mathrm{out} + \Omega_m$ and, obviously, $\delta = \Omega_\mathrm{ref} - \delta_\mathrm{out} - \Omega_m$. As discussed, Eq.~\eqref{eq:OmegaRef} is a well justified estimation of $\Omega_L$ and one can actually use this equation to take a fairly accurate value for the latter.
    
    This same approach was first presented in Ref.~\cite{Nousios2023} but  in the absence of SA which led to a quite simpler equation for $\omega_\mathrm{ref}$ and, importantly, to an expression that did not depend on $\bar n_\mathrm{SA}$. The latter is a consequence of the loss saturation, showing that the exact lasing frequency depends on the ohmic loss level and this is accounted for here. Numerically, though, $\bar n_\mathrm{SA}$ is also unknown and, strictly speaking, Eq.~\eqref{eq:OmegaRef} is not very useful. However, one can either solve it iteratively until convergence [by appointing a value to $\bar n_\mathrm{SA}$, finding $\Omega_\mathrm{ref}$, solving Eq.~\eqref{eq:CMTdynSAApp}, and recalculating] or start from a small pumping rate at which $\bar n_\mathrm{SA} \rightarrow 0$ and use the calculated $\bar n_\mathrm{SA}$ as constant for the next (ascending) step of the calculation. We have verified that both approaches make Eq.~\eqref{eq:g1Relation} perform quite well (obviously in CW), thus leading to fairly good elimination of $\tilde u$ oscillations and, ultimately, to an accurate estimation of $\omega_L$. On the contrary, for pulsed operation, Eq.~\eqref{eq:OmegaRef} can only be evaluated using any abstraction on the choice of $\bar n_\mathrm{SA}$ (e.g., its mean value over the duration of the pulse), but any such option cannot obviously eliminate the oscillations on $\tilde u$. Therefore, and for pulsed operation, Eq.~\eqref{eq:OmegaRef} is as good as Eq.~(25) of Ref.~\cite{Nousios2023}, or frankly any other choice of $\omega_\mathrm{ref}$ reasonably close to $\omega_c$.
    
    Finally, we should note that both $g_1$ and $g_2$ depend on $\omega_\mathrm{ref}$ [see Eq.~\eqref{eq:NormalQuant}]. Thus, their value should be evaluated \emph{after} fixing $\delta$ and $\omega_\mathrm{ref}$. Combining Eq.~\eqref{eq:NormalQuant} with Eq.~\eqref{eq:g1Relation}, we obtain
    \begin{subequations}
        \begin{align}
            \mathrm{Re}\{g_1\} &= \mathcal{A}\frac{\Omega_\mathrm{ref}^2G_m}{\left(\Omega_m^2-\Omega_\mathrm{ref}^2\right)^2+\Omega_\mathrm{ref}^2G_m^2}, \\
            \mathrm{Im}\{g_1\} &= \frac{1}{1+r_\mathrm{SA}\left(1-\displaystyle\frac{\bar n_\mathrm{SA}}{2}\right)}\delta\mathcal{A}\frac{\Omega_\mathrm{ref}^2G_m}{\left(\Omega_m^2-\Omega_\mathrm{ref}^2\right)^2+\Omega_\mathrm{ref}^2G_m^2}, \\
            g_2 &= \mathcal{B}\mathrm{Re}\left\{\frac{j\Omega_\mathrm{ref}}{\Omega_m^2-\Omega_\mathrm{ref}^2+j\Omega_\mathrm{ref}G_m}\right\},
        \end{align}
        \label{eq:g1g2Relation}
    \end{subequations}
    with $\mathcal{A} = \xi_1\sigma_m\tau_\ell^2 N_\mathrm{tot}$ and $\mathcal{B} = (\xi_2/2\hbar\omega_m)\sigma_m\tau_\ell^2 W_\mathrm{sat}$, two parameters that only depend 
    on physical properties of the underlying structure.

\section{\label{app:AppC}Comparison between the CMT framework and the Yamada model}
\renewcommand{\theequation}{C\arabic{equation}}
\setcounter{equation}{0}

    \blue{The CMT model used throughout this work is not a phenomenological model; it is a powerful tool derived from the semi-classical Maxwell-Bloch equations that rigorously describes the physical system under study be means of calculated coefficients. It can be used to model not only lasing in nanophotonic cavities but simultaneously other nonlinear effects as well, e.g., Kerr nonlinearity or saturable absorption that is demonstrated here \cite{Nousios2022,Christopoulos2024Tut}. 
    When only a specific set of phenomena is to be considered, simplified models might be equally useful. For $Q$-switching in semiconductor lasers, for example, there exists the \emph{Yamada model} \cite{Yamada1993,Dubbeldam1999}. We will next show how one can reduce the CMT ODEs to the Yamada equations and under which simplifying assumptions.}
    
    \blue{The Yamada model is a set of three ODEs: The first describes the evolution of the photon number, $S$, the second  the excited electron density in the gain medium, $N_1$, and the third the excited electron density in the SA medium, $N_2$. They are expressed as \cite{Yamada1993,Dubbeldam1999}}
    \blue{\begin{subequations}
            \begin{align}
                \fullFRAC{S}{t} &= [g_1^Y(N_1-N_{t1}^Y)+g_2^Y(N_2-N_{t2}^Y)-\Gamma_0^Y]S + C^Y\frac{N_1}{\tau_s^Y}, \\
                \fullFRAC{N_1}{t} &= J_p^Y - \frac{N_1}{\tau_{s1}^Y} - g_1^Y(N_1-N_{t1}^Y)S - \frac{N_1-N_2}{\tau_d^Y}, \\
                \fullFRAC{N_2}{t} &= - \frac{N_2}{\tau_{s2}^Y} - g_2^Y(N_2-N_{t2}^Y)S - \frac{N_2-N_1}{\tau_d^Y},
            \end{align}
            \label{eq:YamadaFULL}
    \end{subequations}}
    \blue{and share many similarities with our CMT equations [Eqs.~\eqref{eq:CMTinitial},~\eqref{eq:DeltaNinitial},~\eqref{eq:NSAinitial}]. In Eq.~\eqref{eq:YamadaFULL}, $g_i^Y$ ($i=\{1,2\}$) are the differential gain and absorption coefficients, respectively, $N_{ti}^Y$ the transparency values of carrier density for gain and absorption, respectively, $\Gamma_0^Y$ the non-saturable losses, $C^Y$ the spontaneous emission coefficient, $J_p^Y$ the pump current, $\tau_{si}^Y$ the carrier lifetime for the gain and SA media, respectively, and $\tau_d^Y$ an effective lifetime to capture carrier diffusion between gain and SA media. Note the use of the superscript ``$Y$'' here to denote parameters of the Yamada model.}

    \blue{Let us now discuss the constrains of Eqs.~\eqref{eq:YamadaFULL}. As presented in Ref.~\cite{Yamada1993}, they are accurate when the carrier distribution in gain and SA media is uniform;  consequently, the same is assumed for the supported mode $E$-distribution. Additionally, as Eq.~\eqref{eq:YamadaFULL}(a) is cast in terms of number of photons (real quantity), phase-related interactions are not captured, indirectly precluding the introduction of detuning or any interference phenomenon with other waves. 
    Moreover, the coefficients $g_i^Y$ can be considered as a simple phenomenological description of the overlap factors $\xi_i^Y$, which in our case rigorously capture the spatial overlap of the supported lasing mode with the materials. Finally, in the original model the lifetime of gain and SA media was assumed equal ($\tau_{s1}=\tau_{s2}=\tau_s$), but this limitation can be easily lifted \cite{Otupiri2020}.}
    
    \blue{To allow for a direct comparison of Eqs.~\eqref{eq:YamadaFULL} with our CMT framework  we have to set $C^Y=0$ and $\tau_d^Y\rightarrow\infty$, since we have not considered spontaneous emission nor carrier diffusion between different media. Then, we can write}
    \blue{\begin{subequations}
            \begin{align}
                \fullFRAC{S}{t} &= [g_1^Y(N_1-N_{t1}^Y)+g_2^Y(N_2-N_{t2}^Y)-\Gamma_0^Y]S, \label{eq:YamadaPhot} \\
                \fullFRAC{N_1}{t} &= J_p^Y - \frac{N_1}{\tau_{s1}^Y} - g_1^Y(N_1-N_{t1}^Y)S, \label{eq:YamadaNgain} \\
                \fullFRAC{N_2}{t} &= - \frac{N_2}{\tau_{s2}^Y} - g_2^Y(N_2-N_{t2}^Y)S, \label{eq:YamadaNSA}
            \end{align}
            \label{eq:Yamada}
    \end{subequations}}

    \blue{Let us now return to our coupled-mode theory ODEs [Eqs.~\eqref{eq:CMTinitial},~\eqref{eq:DeltaNinitial},~\eqref{eq:NSAinitial}]. We need to relate the quantities of the Yamada model to those in our CMT one. For the carrier densities, the connection is rather obvious: $\bar N_\mathrm{SA} \equiv N_2$ and $\Delta\bar N \equiv N_1-N_{t1}^Y$. Note, however, that our CMT quantities are spatially weighted with the field distribution \cite{Chua2011,Nousios2023,Christopoulos2024Tut} rather than assumed to be spatially uniform. Finally, to relate the cavity amplitude $\tilde a$ with the photon number $S$, one shall recall that the former is normalized so that $|\tilde a|^2$ is equal to the stored energy in the cavity. Since the emitted photon energy equals $\hbar\omega_m$, one can relate the two quantities as $|\tilde a|^2\equiv\hbar\omega_m S$. We stress again that the phase information related with the complex amplitude $\tilde a$ is lost. In summary, we can relate the three quantities as}
    \blue{\begin{subequations}
            \begin{align}
                |\tilde a|^2 &\equiv \hbar\omega_m S, \label{eq:YamadaEquivalenceAmpls} \\
                \Delta\bar N &\equiv N_1-N_{t1}^Y, \label{eq:YamadaEquivalenceNgain} \\
                \bar N_\mathrm{SA} &\equiv N_2. \label{eq:YamadaEquivalenceNSA}
            \end{align}
            \label{eq:YamadaEquivalence}
    \end{subequations}}

    \blue{Considering the introduced equivalences in Eqs.~\eqref{eq:YamadaEquivalence}, one can recast Eqs.~\eqref{eq:CMTinitial},~\eqref{eq:DeltaNinitial},~\eqref{eq:NSAinitial} in terms of $S$, $N_1$, and $N_2$. Equation~\eqref{eq:CMTinitial} should first be expressed in terms of $|\tilde a|^2$ utilizing the chain rule $\mathrm{d}|\tilde a|^2/\mathrm{d}t = \tilde a(\mathrm{d}\tilde a^*/\mathrm{d}t) + \tilde a^*(\mathrm{d}\tilde a/\mathrm{d}t)$, with $\tilde a^*$ denoting the complex conjugate of $\tilde a$. Then, we get}
    \begin{subequations}
            \begin{align}
                \fullFRAC{S}{t} &= -\frac{2}{\tau_\ell}S + 2\xi_1\mathrm{Re}\left\{\frac{j\omega_\mathrm{ref}\sigma_m}{\omega_m^2-\omega_\mathrm{ref}^2+j\omega_\mathrm{ref}\Gamma_m}\right\}(N_1-N_{t1}^Y) S + \frac{1}{N_\mathrm{sat}\tau_\mathrm{SA,0}}(N_2-2N_\mathrm{sat})S, \label{eq:CMTtoYamadaAmpl}\\
                \fullFRAC{N_1}{t} &= \left(R_p+\frac{N_{t1}^Y}{\tau_{21}}\right) - \frac{N_1}{\tau_{21}} - \frac{\xi_2}{2\hbar\omega_m}\mathrm{Re}\left\{\frac{j\omega_\mathrm{ref}\sigma_m}{\omega_m^2-\omega_\mathrm{ref}^2+j\omega_\mathrm{ref}\Gamma_m}\right\}(N_1-N_{t1}^Y)S, \label{eq:CMTtoYamadaNgain}\\
                \fullFRAC{N_2}{t} &= -\frac{\xi_3}{2N_\mathrm{sat}}(N_2-2N_\mathrm{sat})S - \frac{N_2}{\tau_\mathrm{SA}}. \label{eq:CMTtoYamadanSA}
            \end{align}
            \label{eq:CMTtoYamada}
    \end{subequations}  
    
    \blue{It is now straightforward to compare Eqs.~\eqref{eq:CMTtoYamada} with Eqs.~\eqref{eq:Yamada} and retrieve the corresponding coefficients. These are:}
    \blue{\begin{subequations}
            \begin{align}
                \Gamma_0^Y &= \frac{2}{\tau_\ell},\\
                g_1^Y &= 2\xi_1\mathrm{Re}\left\{\frac{j\omega_\mathrm{ref}\sigma_m}{\omega_m^2-\omega_\mathrm{ref}^2+j\omega_\mathrm{ref}\Gamma_m}\right\} \nonumber \\ 
                &= \frac{\xi_2}{2}\mathrm{Re}\left\{\frac{j\omega_\mathrm{ref}\sigma_m}{\omega_m^2-\omega_\mathrm{ref}^2+j\omega_\mathrm{ref}\Gamma_m}\right\}, \label{eq:YamadaEquivalenceCoeffs2} \\
                g_2^Y &= \frac{1}{N_\mathrm{sat}\tau_\mathrm{SA,0}} = \frac{\xi_3\hbar\omega_m}{2N_\mathrm{sat}}, \label{eq:YamadaEquivalenceCoeffs3} \\
                J_p^Y &= R_p+\frac{N_{t1}^Y}{\tau_{21}}, \\
                \tau_{s1}^Y &= \tau_{21}, \\
                \tau_{s2}^Y &= \tau_\mathrm{SA}, \\
                N_{2t}^Y &= 2N_\mathrm{sat}.
            \end{align}
            \label{eq:YamadaEquivalenceCoeffs}
        \end{subequations}}
    \blue{Equations~\eqref{eq:YamadaEquivalenceCoeffs2}~and~\eqref{eq:YamadaEquivalenceCoeffs3} further signify that our CMT framework is more general than the Yamada model. Specifically, in Eq.~\eqref{eq:YamadaEquivalenceCoeffs2} it is assumed that $\xi_2 = 4\xi_1$, an equality that is strictly true only if the field distribution is uniform inside the laser cavity \cite{Nousios2023}. Similarly for Eq.~\eqref{eq:YamadaEquivalenceCoeffs3}, it is implied that $\xi_3 = 2/(\hbar\omega_m\tau_\mathrm{SA,0})$, which is again  only true when the field distribution is uniform in the laser cavity \cite{Ataloglou2018,Nousios2022}.}

    \blue{Before concluding, we would like to further elaborate on the main advantage of our framework compared to the Yamada model, i.e., the complex-valued cavity amplitude instead of the real-valued number of emitted photons inside the cavity. In the context of this work, this is only reflected in the detuning parameter, which is absent from Eqs.~\eqref{eq:YamadaFULL}. Its inclusion allows for the demonstration of $Q$-switching with additional 2D materials at the expense of higher pumping (see Sect.~\ref{subsec:ExamplDet}); the Yamada model cannot capture such an effect. Moreover, in more complex systems, e.g., in systems with multiple cavities \cite{Yacomotti2013,Rasmussen2017}, the correct consideration of phase is crucial for capturing interference effects.}

    \blue{Let us further elaborate on this last point by introducing the notation $\tilde u(t) = |\tilde u(t)|\exp\{j\varphi_u(t)\}$ in Eq.~\eqref{eq:CMTdynSAAmplApp}, with $|\tilde u(t)|$ and $\varphi_u(t)$ being the time-dependent amplitude and phase of $\tilde u(t)$ (both real-valued quantities). We use the normalized instead of the full version of our CMT framework solely to avoid lengthy notations. It is then not hard to obtain}
    \blue{\begin{subequations}
            \begin{align}
                \fullFRAC{|\tilde u|}{t^\prime} &= -|\tilde u| + \mathrm{Re}\{g_1(\delta)\}\Delta\bar n|\tilde u| - r_\mathrm{SA}\left( 1 - \frac{\bar n_\mathrm{SA}}{2} \right)|\tilde u|, \label{eq:CMTdynSAAmplDisentagledSmpl} \\
                \fullFRAC{\varphi_u}{t^\prime} &= -\delta + \mathrm{Im}\{g_1(\delta)\}\Delta\bar n. \label{eq:CMTdynSAAmplDisentagledPhase}
            \end{align}
            \label{eq:CMTdynSAAmplDisentagled}
        \end{subequations}}
    \blue{Clearly, detuning is solely related to the gain process, and a $\delta\neq 0$ is only relevant when the cavity resonance frequency, $\omega_c$, does not coincide with the peak emission frequency of the gain medium, $\omega_m$, i.e., for $\delta_\mathrm{out}\neq 0$. Such a term is missing from the Yamada model. To further indicate its importance, let us consider the following instructive example: Let us assume that $\omega_c\neq\omega_m$ ($\delta_\mathrm{out}\neq 0$) but, instead of following the process of Appendix~\ref{app:AppΒ}, we choose $\omega_\mathrm{ref} = \omega_c$ so that $\delta=0$ [cf. Eq.~\eqref{eq:NormalQuantDet}]. Such a choice superficially gives the impression that the phase contribution is absent from Eq.~\eqref{eq:CMTdynSAAmplApp}. However, this is not true, as revealed from Eqs.~\eqref{eq:CMTdynSAAmplDisentagledPhase}~and~\eqref{eq:NormalQuantg1} since $\mathrm{Im}\{g_1\}\neq 0$. On the other hand, fixing $g_1$ to a real value implies the choice $\omega_\mathrm{ref} = \omega_m$, which unavoidably results in $\delta\neq 0$ and again introduces a phase contribution and a non-trivial solution of Eq.~\eqref{eq:CMTdynSAAmplDisentagledPhase}. In fact, any choice of $\omega_\mathrm{ref}$ (or $\delta$) is equally valid, as long as the involved parameters are consistently calculated, and introduces an associated phase term through Eq.~\eqref{eq:CMTdynSAAmplDisentagledPhase}. The approach that we followed for the calculations in this work is the one described in Appendix~\ref{app:AppΒ} and it is optimum only in terms of numerical efficiency.}

\bibliographystyle{ieeetr}
\bibliography{Bibliography}
\end{document}